\documentclass[12pt]{iopart} 
\usepackage{iopams}  
\usepackage[english]{babel}
\usepackage{natbib}
\usepackage[utf8]{inputenc}
\usepackage[margin=0.75in]{geometry}
\usepackage{multirow}
\usepackage{subcaption}
\usepackage{graphicx}
\usepackage{xcolor}
\begin{document}

\title[Investigating the robustness of a learning-based method for quantitative phase retrieval]{Investigating the robustness of a learning-based method for quantitative phase retrieval from propagation-based x-ray phase contrast measurements under laboratory conditions}

\author{Rucha Deshpande$^1$, Ashish Avachat$^{2,3}$, Frank J. Brooks$^3$, \\ Mark A. Anastasio$^3$}
\address{$^1$ Department of Biomedical Engineering, Washington University in St. Louis, St. Louis, MO, USA.}
\address{$^2$ Department of Mechanical Engineering and Materials Science, University of Pittsburgh, Pittsburgh, PA, USA.}
\address{$^3$ Grainger College of Engineering, Department of Bioengineering, University of Illinois Urbana-Champaign, Urbana, IL, USA.}
\ead{maa@illinois.edu}

\section*{Abstract}
\textit{Objective.} Quantitative phase retrieval (QPR) in propagation-based x-ray phase contrast imaging of heterogeneous and structurally complicated objects is challenging under laboratory conditions due to partial spatial coherence and polychromaticity. A learning-based method (LBM) provides a non-linear approach to this problem while not being constrained by restrictive assumptions about object properties and beam coherence. The objective of this work is to assess a LBM for its applicability under practical scenarios by evaluating its robustness and generalizability under typical experimental variations. \textit{Approach.} Towards this end, an end-to-end LBM was employed for QPR under laboratory conditions and its robustness was investigated across various system and object conditions. The robustness of the method was tested via varying propagation distances and its generalizability with respect to object structure and experimental data was also tested. \textit{Main results.} Although the end-to-end LBM was stable under the studied variations, its successful deployment was found to be affected by choices pertaining to data pre-processing, network training considerations and system modeling. \textit{Significance.} To our knowledge, we demonstrated for the first time, the potential applicability of an end-to-end learning-based quantitative phase retrieval method, trained on simulated data, to experimental propagation-based x-ray phase contrast measurements acquired under laboratory conditions with a commercial x-ray source and a conventional detector. We considered conditions of polychromaticity, partial spatial coherence, and high noise levels, typical to laboratory conditions. This work further explored the robustness of this method to practical variations in propagation distances and object structure with the goal of assessing its potential for experimental use. Such an exploration of any LBM (irrespective of its network architecture) before practical deployment provides an understanding of its potential behavior under experimental settings.   

\noindent{\it Keywords\/}: quantitative phase retrieval, learning-based method, propagation-based x-ray phase contrast imaging, laboratory conditions

\section{Introduction}
X-ray phase-contrast (XPC) imaging enables the acquisition of information about the spatial distribution of the complex valued refractive index of an object, beyond the attenuation contrast provided by conventional x-ray imaging \citep{snigirev1995possibilities}. As a result, XPC imaging has been employed in fields ranging from material science \citep{stevenson2003phase, mayo2012line} to biomedical science \citep{bravin2012x}, where attenuation contrast might be insufficient for resolving certain object features. Furthermore, the advent of polychromatic sources with high spatial coherence further expanded the applicability of XPC imaging to laboratory conditions \citep{wilkins1996phase, hemberg2003liquid, wilkins2014evolution}. Various imaging setups, such as, interferometry-based \citep{bonse1965x}, grating-based \citep{pfeiffer2012milestones}, analyzer-based \citep{chapman1997diffraction}, and propagation-based (PB-XPC) \citep{snigirev1995possibilities, wilkins1996phase} have been employed for XPC imaging, of which the last is particularly simple in terms of the experimental setup. However, images acquired under laboratory conditions are degraded by partial spatial and temporal coherence, causing some loss in image quality as compared to imaging via synchrotron sources. Thus, any method for the recovery of phase effects from mixed-contrast data acquired in laboratory conditions must compensate for partial coherence effects and be stable under typical system variations.

Several methods for QPR from PB-XPC measurements have been proposed over the last few decades \citep{langer2008quantitative, burvall2011phase, haggmark2017comparison, luu2011quantitative, davidoiu2014non}. These methods differ in the estimated quantity (phase \citep{paganin1998noninterferometric, guigay2007mixed} or projected thickness \citep{paganin2002simultaneous, beltran20102d}), source spectrum (monochromatic \citep{haggmark2017comparison, mohan2020constrained} or polychromatic \citep{luu2011quantitative,gursoy2013single}), number of measurements (one, two, or more) \citep{langer2008quantitative}, assumptions about the object (single material \citep{paganin2002simultaneous}, weak absorption \citep{arhatari2008phase}, slowly varying phase \citep{arhatari2008phase, luu2011quantitative}, etc.) or system linearity \citep{gureyev2006linear, davidoiu2013nonlinear}. A learning-based solution to QPR from PB-XPC measurements could be a new approach that allows the circumvention of assumptions about the object, and system linearity. More recently, LBMs have been employed as a solution to the phase retrieval problem in the holographic regime \citep{zeng2021deep}, far-field regime for ptychographic imaging \citep{cherukara2020ai, harder2021deep}, and under monochromatic conditions in the near-field regime \citep{wu2022enhanced, mom2022mixed}. However, their applicability to experimental data acquired via PB-XPC imaging under laboratory conditions has not been shown. Most importantly, before employing a LBM under actual experimental conditions, it is essential to first demonstrate its stability under typical system conditions and material variations. Second, it is equally important to test the generalizability or tolerance of the method towards practical, experimental variations or imprecision. Towards the goal of evaluating the robustness of a LBM under practical, laboratory conditions, we first employed an end-to-end, LBM as a non-linear approach to the QPR problem under partial spatial coherence and polychromaticity conditions; this method is also applicable to complex, multi-material objects manifesting mixed contrast. We then evaluated the performance of this method under varying system and object conditions and finally, quantified its generalizability towards previously unencountered variations in object structure, system settings, and experimental data.

\section{Background}
\subsection{The phase retrieval problem}
The QPR problem for XPC imaging is to estimate the phase induced by the presence of an object in the path of an incident x-ray wavefield from intensity measurements alone. Under laboratory conditions, the presence of partial temporal, and spatial coherence worsens the quality of these measurements. Although propagation-based XPC imaging is robust to imperfect temporal coherence  \citep{wilkins1996phase, paganin2006coherent}, its tolerance to spatial incoherence is significantly worse \citep{paganin2006coherent}. 

In the presented study, we seek to estimate the projected phase that is defined by the imaginary component of the wavefield on the contact plane corresponding to a \emph{single energy} alone from intensity measurements of a mixed-contrast object, acquired under laboratory conditions (considering a polychromatic source spectrum). Typically, for estimation of the wavefield phase when a mixed-contrast object is considered, at least two measurements that contain complementary information are required \citep{gureyev1998x, gureyev2000quantitative, burvall2011phase}. This may be achieved by employing spectral detectors, spectrum switching, multiple monochromatic measurements or measurements at multiple distances \citep{gureyev1998x}. Alternatively, some solutions relate the attenuation and phase effects via the object thickness \citep{paganin2002simultaneous, beltran20102d} to estimate phase from a single measurement. In the present work, two measurements were acquired by varying propagation distances and using a conventional, integrating detector \citep{gureyev1998x, paganin1998noninterferometric}.

\subsection{Canonical measurement model}
Consider a thin, multi-material object located at a distance $R_1$ from the source that possesses compact support, illuminated by a monochromatic plane wave $U(\textbf{r}, \omega)$ propagating along the optical axis $z$. 
The object is characterized by its energy-dependent, complex-valued refractive index given by:

\begin{equation}
    n(\textbf{r},z,\omega) = 1 - \delta(\textbf{r},\omega,z) + i\beta(\textbf{r},\omega,z),
\end{equation}
where $\textbf{r}$ represents a position in two-dimensional space corresponding to the plane transverse to the optical axis, $\omega$ represents the temporal frequency, $\delta$ and $\beta$ respectively represent the real and imaginary parts of the complex refractive index, and $i\equiv\sqrt{-1}$.

Variations in the phase and amplitude of the forward propagating $(z>0)$, monochromatic wavefield $U(\textbf{r},\omega)$ are induced after its interaction with the object. The phase shift induced by the object can be represented as:
\begin{equation}
    \phi(\textbf{r},\omega) = -k\int\delta(\textbf{r},\omega,z)dz,
\end{equation} 
where $k=2\pi/\lambda$ is the wavenumber corresponding to $\omega$.
Similarly, the amplitude modulus $M(\textbf{r}, \omega) = \exp[-A(\textbf{r}, \omega)]$, which represents the attenuation contrast is impacted as:
\begin{equation}
    A(\textbf{r},\omega) = k\int\beta(\textbf{r},\omega,z)dz.
\end{equation}
Thus, the transmitted wavefield $U(\textbf{r},\omega,0)$ immediately behind the object ($z=0$) is given by:
\begin{equation}
   U(\textbf{r}, \omega,0) = \exp[i\phi(\textbf{r}, \omega)-A(\textbf{r}, \omega)]U(\textbf{r},\omega) = T(\textbf{r},\omega)U(\textbf{r},\omega),
\end{equation}
where $T(\textbf{r},\omega)$ is the transmission function. The intensity distribution corresponding to the transmitted wavefield represents the contact plane image: 
\begin{equation}
    I(\textbf{r},\omega,0) = |U(\textbf{r},\omega,0)|^2.
\end{equation}

On propagation through free space, the perturbations in the transmitted wavefield modulate the amplitude and phase of the wavefield. The wavefield at the downstream detector plane ($z=R_2$) can be expressed as:
\begin{equation}
    U(\textbf{r},\omega,R_2) = H(\omega,R_2)*U(\textbf{r},\omega,0),
\end{equation}
where $H$ represents the Fresnel propagator \citep{paganin2006coherent} and * represents the two-dimensional convolution operation. The corresponding intensity distribution is given  $I(\textbf{r},\omega,R_2) = |U(\textbf{r},\omega,R_2)|^2$.

Considering a polychromatic spectrum and illumination by a spherical wave, the intensity on the downstream detector plane can be expressed as \citep{gureyev1998x, paganin2006coherent}: 
\begin{equation}
    I_{poly}(\textbf{r}/M, R_2/M) = M^2 \int_\omega d\omega S(\omega) D(\omega) I(\textbf{r},\omega, R_2), 
\end{equation}
 where $D(\omega)$ is the frequency-dependent efficiency of the detector response, $S(\omega)$ is the spectral weight due to the source spectrum, and  $M = R_1 + R_2 / R_1 $ denotes the magnification. 

The overall system blur resulting from the source and detector characteristics can be modeled as a convolution with the system point spread function (PSF). Additionally, measurement noise modeled as a mixed Poisson-Gaussian random variable. Thus, the final measured image is given by:
\begin{equation}
    I_{poly}(\textbf{r}/M, R_2/M, \sigma) = M^2 \int_\omega d\omega S(\omega) D(\omega) I(\textbf{r},\omega, R_2) * PSF(\textbf{r},\sigma) + \eta,
\end{equation}
where $\sigma$ denotes the width of the system blur, $PSF(\textbf{r},\sigma)$ denotes the system point-spread-function and $\eta$ is the additive system noise.

\subsection{Related work}
\subsubsection{Conventional methods for phase retrieval in PB-XPC}
Many popular phase retrieval methods from PB-XPC measurements involve a linear solution to the phase retrieval problem under monochromatic source conditions \citep{paganin2002simultaneous, langer2008quantitative, burvall2011phase}. These methods are typically based on the transport-of-intensity equation (TIE) formulation \citep{teague1983deterministic} or the contrast transfer function (CTF) formulation \citep{pogany1997contrast} and may employ one \citep{paganin2002simultaneous, paganin2020boosting}, two \citep{paganin1998noninterferometric}, or more \citep{guigay2007mixed} measurements. While such deterministic methods are simple to implement and physically intuitive, they require restrictive assumptions such as a single material object, weak absorption, or slowly varying phase. Alternatively, iterative methods (also under monochromatic conditions) have been proposed in the linear formulation \citep{luu2011quantitative, yan2010performance} and non-linear formulations \citep{davidoiu2014non, davidoiu2012nonlinear, mohan2020constrained}. However, these methods generally have higher sampling requirements than analytic methods. 
Phase retrieval from polychromatic measurements, typical to laboratory conditions, has involved (i) extension of deterministic formulations \citep{gureyev1998x, lohr2020comparison} that are subject to assumptions about objects or system settings or (ii) iterative methods, demonstrated on very thin ($\sim$1 $\mu$m) objects, and with a linearized system model \citep{carroll2017iterative} of the inherently non-linear mapping between intensity and phase. Thus, there is a need for more widely applicable QPR methods for use with PB-XPC imaging that are suitable for experimental imaging of heterogeneous objects under laboratory conditions and don't rely on assumptions about object properties or system linearity. Towards this goal, LBMs represent potential solutions that addresses these issues.

\subsubsection{LBMs for phase retrieval}
Prior studies of LBMs for phase retrieval in optical imaging in the holographic regime have shown promise \citep{zeng2021deep, wijesinghe2021emergent, cherukara2018real, deng2020probing, kang2020phase, li2021patch, li2021u, luo2021cascaded, zhang2020holo}. Recently, a few methods have also explored their applicability for XPC imaging in the near-field regime \citep{wu2022enhanced, mom2022mixed, zhang2021phasegan, xu2022single}. However, most LBMs for phase retrieval from XPC measurements have been proposed for \emph{monochromatic} conditions. Although two of these recent studies \citep{wu2022enhanced, xu2022single} assume a laboratory setup, the first study \citep{wu2022enhanced} considers a single effective energy for numerical studies, while the second \citep{xu2022single} involves a grating-based imaging setup. In the present work, we employ and assess a LBM for (i) QPR from PB-XPC measurements acquired under laboratory conditions with consideration of multi-material objects, (ii) robustness to commonly encountered practical variations in system parameters and object structure and (iii) generalizability to experimental data. Thus, this work aims to investigate the practical applicability and limitations of employing an LBM for QPR from heterogeneous and structurally complicated objects imaged via PB-XPC under laboratory conditions.

\section{Methods}
\subsection{Overview of approach and study goals}

We propose a learning-based solution to the QPR problem under conditions of polychromaticity and partial coherence, typical to laboratory conditions. This approach circumvents the assumptions of the existing methods by learning an analytically intractable, non-linear mapping from the acquired intensity measurements to the object-induced wavefield phase. A LBM seeks to learn this mapping from a training dataset constituting sufficient representations or instances of this mapping. Each instance of this mapping consists of (i) two input intensity measurements --- one acquired at the contact plane and the second at a downstream detector plane and (ii) a target phase map that represents the wavefield phase on the contact plane corresponding to the peak energy of the spectrum. 
The rationale for this setup is that at least two inputs are generally required for unambiguous phase retrieval when a multi-material object provides both attenuation, and phase contrast \citep{gureyev1998x, gureyev2000quantitative, burvall2011phase}. Furthermore, to retrieve phase corresponding to a single energy, from polychromatic data, a network employed as a LBM is expected to parse the effects of the spectral distribution and learn the unique mapping (in a practical sense) between intensity and phase \citep{nugent2007x, burvall2011phase} corresponding to \emph{only the peak energy}. Note that in this study we consider typical practical conditions and the absence of phenomena such as phase vortices and zeros.

The proposed LBM employs an encoder-decoder like U-Net architecture \citep{ronneberger2015u}, which has been shown previously to be highly effective for data-driven solutions for inverse problems \citep{jin2017deep}. However, when employing any LBM, it is not sufficient to merely demonstrate the capability of the method to solve such a problem, but it is also necessary to ensure that the method can be employed stably under variations in system or object conditions that may be practically encountered. Besides stability, considerations of generalizability and dataset sufficiency are also important. Thus, the focus of this work is to investigate the robustness of a learning-based solution for QPR, particularly with regard to practical utility. Specifically, the following properties of a LBM employed for QPR from PB-XPC measurements under laboratory conditions will be explored:
(i) stability under typical laboratory conditions,
(ii) quantitative accuracy,
(iii) qualitative performance related to retention of features of interest,
(iv) robustness to practical variations in propagation distances and object structure.
To enable a systematic study of these issues, computer-simulations were conducted. Additionally, experimental data was utilized to investigate the direct experimental generalizability of the LBM trained on simulated data.
The following sub-sections detail the methodology employed in the studies.

\subsection{Numerical phantom design}

Two classes of phantoms were employed in the computer-simulation studies (i) a phantom comprising uniform spheres and (ii) a complex phantom comprising spheres and ellipsoids.
For the first type, referred to as  ``spheres phantoms", the phantoms were specified by a fixed number of uniform spheres (16, 32 or 64), but of variable radii drawn from a beta distribution (mean: N/16, std dev: 20\%, range: N/32 to N/8), where N is the side of the detector plane in pixels. In all cases, N=1024 pixels or 13.5 mm. In physical units, the mean sphere radius corresponds to approximately 845 $\mu$m. The spheres were placed randomly within the field-of-view (FOV), with at least a distance of N/4+16 pixels away from the detector edges at the contact plane to ensure compact support for the object even in the downstream measured image. This also ensured that the spheres did not intersect each other in 3D although the projected thickness might still show overlaps in 2D as the spheres were separated only along the projection axis. 

The following  configurations of the spheres phantoms were considered: (a) single material phantom: soft-tissue in air, (b) multi-material phantom without any embedding medium: four biological tissues (lung, liver, cartilage, intestine), and (c) multi-material phantom including an embedding material: the embedding material corresponds to muscle, and the refractive indices of the three embedded materials were obtained by increasing material density by 0.5, 1 and 1.5 times with respect to the embedding material (muscle). In case (c), the total object thickness was taken as 13.5 mm, inducing non-negligible attenuation contrast. This value equals the detector side length and thus, describes the object as a cube. Note that object thickness can be varied and this value was chosen merely to demonstrate the applicability of the method in a regime of realistic samples demonstrating phase contrast (few centimetres), going beyond very thin or weak attenuation samples (few microns). The complex-valued refractive indices were computed by use of the \textit{xraylib} library \citep{schoonjans2011xraylib} and elemental compositions of tissues \citep{russo2017handbook}. 

The second class of phantoms, referred to as the ``complex phantoms", provided a greater variation in shapes and sizes of the individual components forming an object than the spheres phantoms, and was designed specifically for the assessment of the generalizability to variations in object structure. Each realization of a complex phantom comprised 64 structures drawn uniformly at random from a distribution of four distinct shapes: small spheres (mean radius: N/16, std. dev.: 0.5$\times$mean), large spheres (mean radius: N/32, std. dev.: 0.5$\times$mean), large ellipsoids (mean semi-axes lengths: N/16, N/8, N/16, std. dev.: 0.5$\times$mean), thin ellipsoids (mean semi-axes lengths: N/64, N/8, N/64, std. dev.: 0.05$\times$mean). Thus, while employing simple, analytically described shapes such as spheres and ellipsoids, this phantom provided greater structural complexity than the spheres phantom due to multiple, overlapping (in projection) constituent structures. The distribution of the constituent shapes in this case is only representative of a complex phantom, and could easily be varied to obtain similar structural complexity. Material allocation for this phantom corresponded to soft-tissue.  

\subsection{System simulation and image formation}
\begin{figure}[h!tb]
\centering
\begin{subfigure}{0.9\textwidth}
    \includegraphics[width=0.9\linewidth]{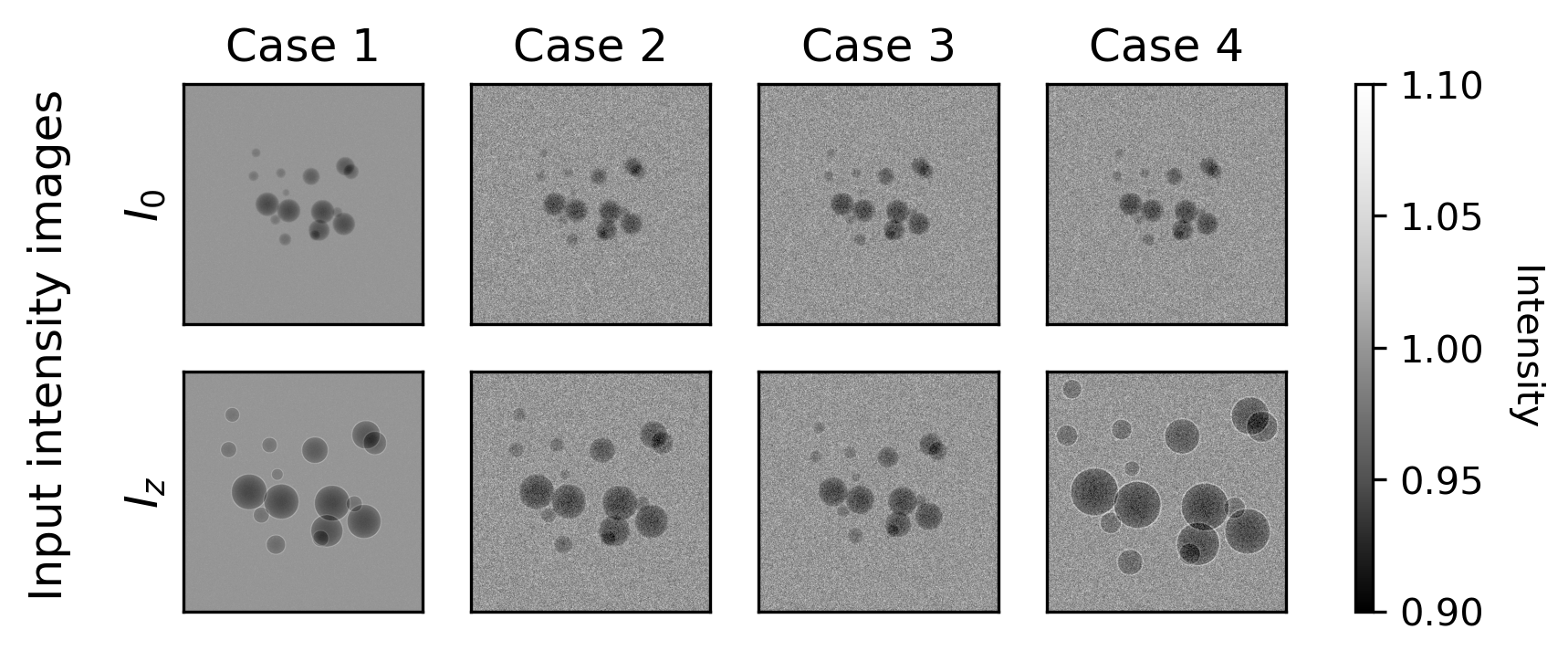} 
\end{subfigure}
\caption{Sample input intensity images under various system conditions. (Top row) Contact plane images. (Bottom row) Intensity images acquired after propagation. This image-pair serves as the two-channel input to a LBM. Case 1: low blur (FWHM: 39 $\mu$m) and low noise (0.5\%). Case 2: high blur (FWHM: 92 $\mu$m) and high noise (5\%). Case 3: low $R_2$ (0.25 m), high noise (5\%). Case 4: high $R_2$ (1.00 m), high noise (5\%).}
\label{fig1}
\end{figure}

For use in network training, simulated PB-XPC image data were produced by imaging the phantoms virtually. The source spectrum was modeled to match an x-ray source with a liquid-metal-jet anode (MetalJet D2, Excillum) operated at 70 kVp, at 110~W power (peak energy of 24 keV and equivalent energy 19 keV), and filtered through 0.9 mm Aluminuium. The spectrum was resampled in steps of 1 keV in the Bremsstrahlung (non-peak) region and 0.2 keV near the peak, yielding 65 discrete energy levels in total. The detector was modeled as a CsI(Tl) scintillator-based X-ray imager with a pixel pitch of 13 $\mu$m. At each of these energies, the virtual image was formed by the interaction of the object with the incident radiation, modeled as a convolution of the object transmission function with the Fresnel propagator \citep{voelz2011computational} in the Fourier domain (refer Sec 2.1). To simulate the effects of a cone beam geometry, the Fresnel scaling theorem \citep{paganin2006coherent} was employed after propagation. For the geometries simulated in this work the source-to-object distance ($R_1$) was set to 1 m and the object-to-detector distance ($R_2$) was varied from 0.25 to 1 m, yielding magnifications in the range 1.25 -- 2. All simulations were oversampled by a factor of 2 to ensure sufficient sampling \citep{voelz2011computational}. The monochromatic images at each discrete energy bin were then combined to form a single polychromatic image according to their spectral weights and the energy-specific detective sensitivity. Next, the effect of system blur was modeled as a convolution of the virtual image with a Gaussian kernel whose width was determined by the full-width-half-maximum (FWHM) of the system blur, ranging from 39.6 to 92.4 $\mu$m, and corresponding to 3, 5 and 7 detector pixels respectively. Last, the image was downsampled to yield a 1024$\times$1024 image and system noise was added, modeled as a mixed Poisson-Gaussian random variable possessing zero mean and standard deviation ranging from 0.5 to 5 \% of individual pixel intensity. 
A contact plane image ($I_0$) and the downstream measured image ($I_z$) form a single input pair to the U-Net, and examples are shown in Figure \ref{fig1}. The corresponding target image that the U-Net seeks to approximate was specified as a phase map of the projected thickness at 24 keV energy, and computed as $\phi_{E=24 keV} = -k\delta_{E=24 keV}t$ where $t$ is the projected thickness of the object. Thus, a single sample from the simulated training dataset consisted of the two detector images $(I_0, I_z$) as inputs and the target phase map as the output.   

\subsection{Network trainings}
A U-Net deep neural network architecture \citep{ronneberger2015u} was employed that contained two input- and one output-channel. All images were of dimension 1024$\times$1024. It should be noted that the FOV for each of the two input images  $(I_0, I_z$) is different, and determined by the magnification. The two input channels were normalized independently over the entire dataset. This ensured that the mapping from intensity to phase is retained while also aiding training stability. Note that while global normalization ensures the retention of the information of this physical mapping, a per-image normalization would break this mapping and render the problem unsolvable. 

A network of depth three was chosen and trained to minimize the MSE loss using an Adam optimizer (learning rate $= 5e-5$ to $5e-6, \beta_1 = 0.9, \beta_2 = 0.999, \epsilon = 1e-7$). Other parameter choices included: kernels of size 3$\times$3 in the encoder/decoder blocks and 1$\times$1 in the skip connections, upsampling via nearest-neighbor interpolation, and downsampling via max-pooling. Model training for the case of objects without an embedding medium (zero background phase value) was performed for 100 epochs with a batch size of 1, and the model with the least validation loss was chosen for inference. For the case of objects embedded within a surrounding medium (non-zero background phase value), training was performed with an adaptive learning rate schedule ($lr = 5e-3$ to $5e-4$) over 160 epochs. Typically, the training, validation and test sets consisted of 2000, 1000 and 1000 images respectively. Model training was performed on Nvidia GeForce GTX TitanX, 1080-Ti and A100 GPUs and typically took 15 to 60 hours per training. 

For the numerical studies of variations in system conditions, materials and phantoms, a separate network was trained for each combination of system and object parameters. For the generalizability studies, a single network trained for a certain object and geometry combination was tested on multiple, out-of-distribution datasets. 

\subsection{Overview of experiments}
The performance of the LBM was assessed under variations in (i) system parameters: blur and noise, and object-to-detector distance ($R_2$), and (ii) object composition: multi-material objects in the presence or absence of an embedding medium. The generalizability of the method was tested on out-of-distribution simulated datasets corresponding to: (i) varying values of $R_2$, and (ii) complexity of object structures. Subsequently, experimental data were also employed. Three measures of performance were used for this purpose: normalized root-mean-square error over the object (NRMSE), structural similarity index measure (SSIM) \citep{wang2004image} and peak signal-to-noise ratio (PSNR). The NRMSE is defined as:
\begin{equation}
   NRMSE = \frac{||\phi_{pred} - \phi_{true} ||_2}{|| \phi_{true}||_2}.
\end{equation}

All comparisons were made against two popular analytic methods for single material objects by Paganin \textit{et al.}\citep{paganin2020boosting}, or multi-material objects embedded in a medium by Beltran \textit{et al.} \citep{beltran20102d}, (here onward referred to as PG and BT respectively). Each of these methods was implemented at two energies: the peak energy of the spectrum --- corresponding to the energy of the target phase map for the LBM, and the equivalent energy of the spectrum. The two variations of the PG/BT methods are referred to as PG/BT-peak and PG/BT-eqvt respectively. These methods are stable under high noise conditions, and have been known to provide greater accuracy and low-frequency stability than the two-plane TIE method \citep{paganin1998noninterferometric, paganin2002simultaneous}. Hence, they are chosen to provide higher benchmarks for comparison. As both methods, PG and BT, recover the object thickness, the phase map was obtained by employing the known values for $k\delta_{(E=24 keV)}$. Furthermore, the multi-material method BT \citep{beltran20102d} requires knowledge of object locations, and this was supplied to the method.
\section{Results}

\subsection{Behaviour of the network under various system conditions}
All phantoms in the system variation studies described below corresponded to sphere phantoms and consisted of a single material, with the refractive index values representative of soft-tissue. This ensured that the impact of system variations were not confounded by material variations, and allowed a fair comparison with conventional methods intended for use with homogeneous objects.

\begin{figure}[h!tb]
\centering
\begin{subfigure}{0.8\textwidth}
    \includegraphics[width=0.9\linewidth]{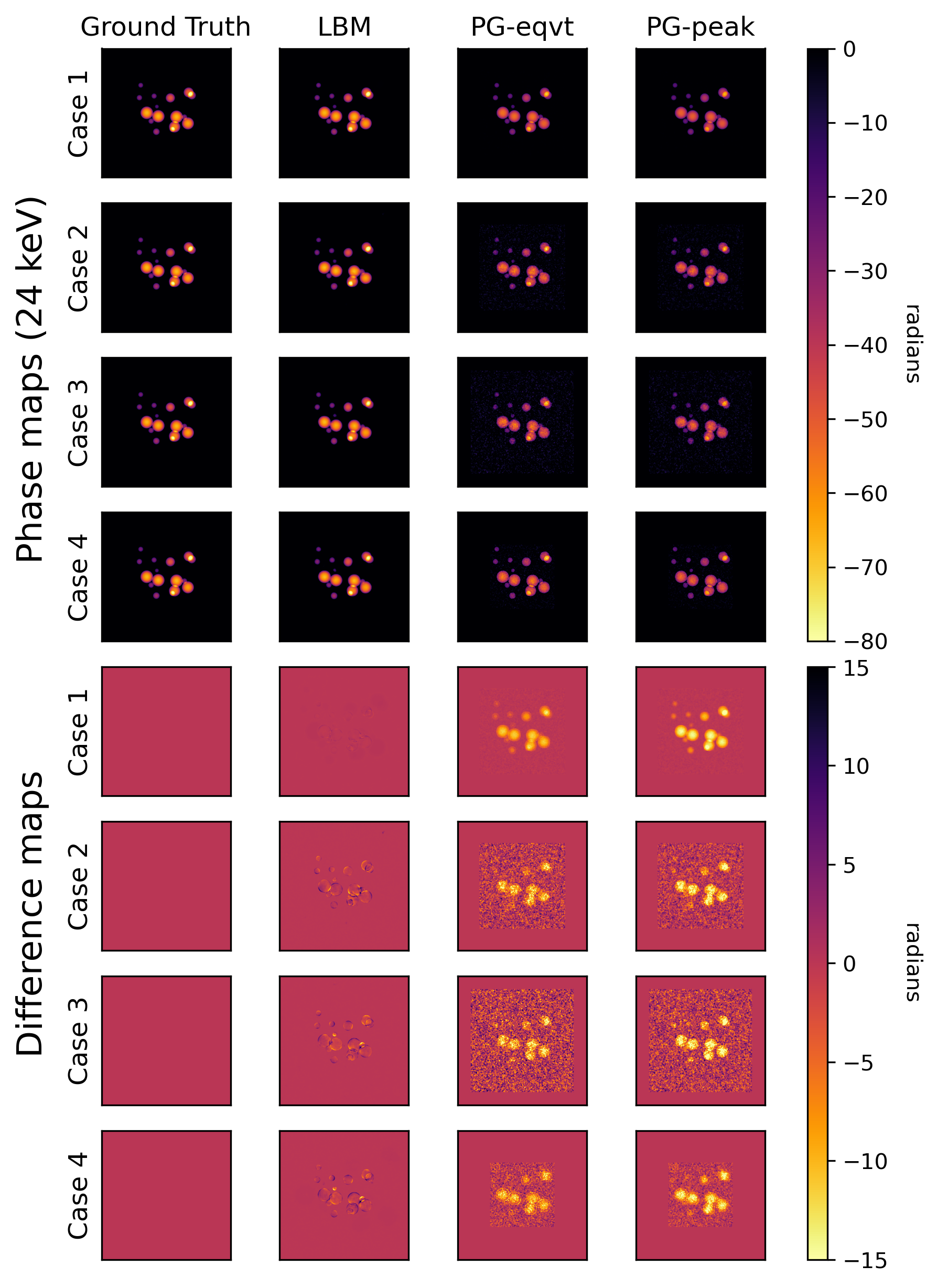} 
\end{subfigure}
\caption{Qualitative results of phase retrieval under various system conditions. Case 1: low blur (FWHM: 39 $\mu$m) and low noise (0.5\%). Case 2: high blur (FWHM: 92 $\mu$m) and high noise (5\%). Case 3: low $R_2$ (0.25 m), high noise (5\%). Case 4: high $R_2$ (1.00 m), high noise (5\%). Performance of the LBM was qualitatively better than that of the conventional methods, even under conditions of high noise and blur as seen in the difference maps. This also holds across varying propagation distances. Note: Predictions from conventional methods are zero padded at the periphery for display. ($R_2$ is fixed at 0.5 m for cases 1 and 2, while blur is fixed at 66 $\mu$m FWHM for cases 3 and 4. Phantom material: soft-tissue)}
\label{fig4}
\end{figure}

\begin{figure}[h!tb]
\centering
\begin{subfigure}{0.6\textwidth}
    \includegraphics[width=0.9\linewidth]{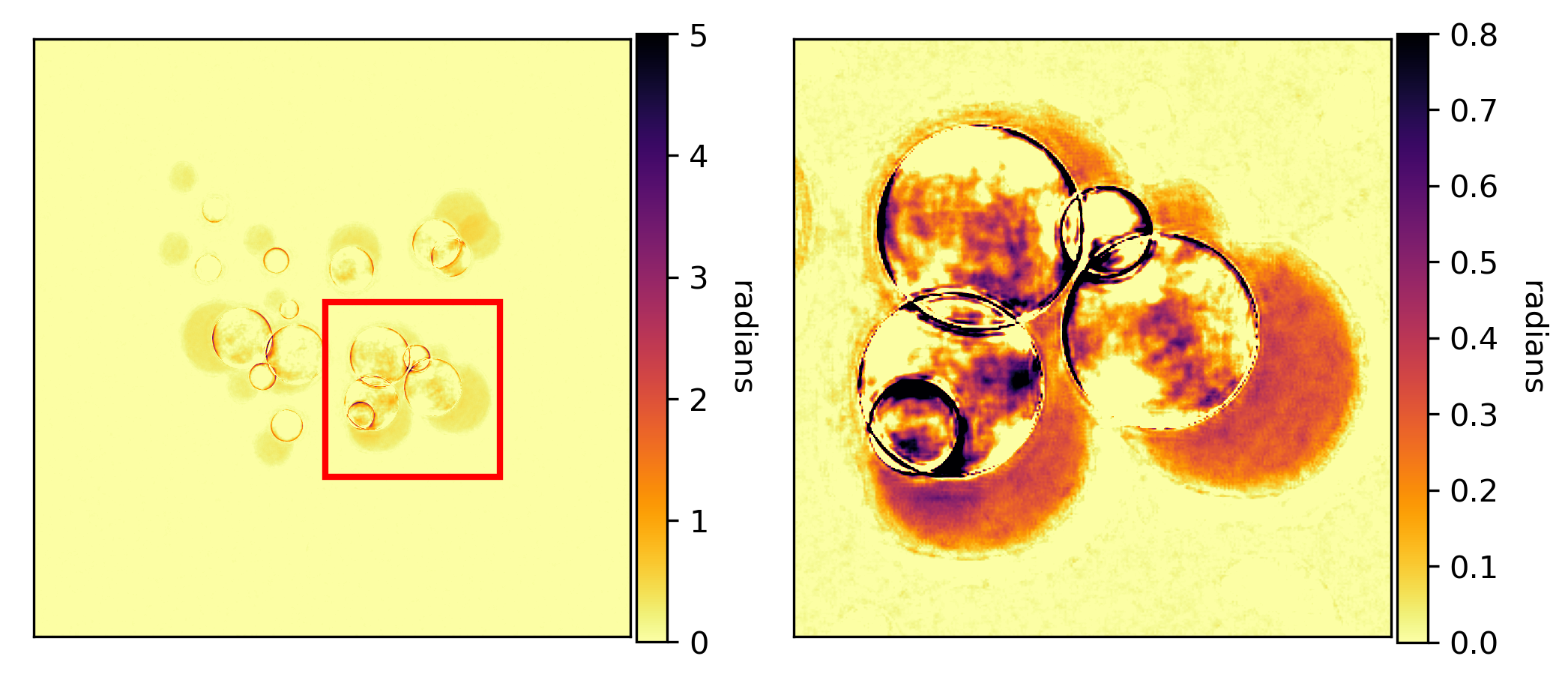} 
\end{subfigure}
\caption{Examples of typical artifacts seen in phase predictions. (Left) Sample difference image shows two types of artifacts: (i) high-frequency artifacts at the object edges and (ii) low-amplitude, residual artifacts corresponding to the object locations in the magnified input image. (Right) Cropped section from the left image shows that the residual error is negligible ($<1$ radian) as compared to the maximum absolute phase in the ground truth ($\approx86$ radians). Additionally, note that the edge artifacts are not uniform in all directions.}
\label{fig6}
\end{figure}

Qualitative results for the studies pertaining to variations in the physical parameters: (i) noise and system blur and (ii) propagation distances, are presented in Figure \ref{fig4}. Generally, the features of the object were well formed although some blur at the edges can be observed in cases with high noise levels (5\%). Predictions from the LBM have noticeably lower noise and, thus, stronger bias as compared to results from the conventional methods under high noise conditions. However, some high-frequency, residual artifacts were observed in the difference images for the LBM predictions. Additionally, very low amplitude artifacts, spatially corresponding to object locations at the magnified FOV were also present (see Figure \ref{fig6}), potentially resulting from the joint tasks of quantitative phase retrieval and demagnification. 

\begin{figure}[h!tb]
\centering
\begin{subfigure}{0.9\textwidth}
    \includegraphics[width=0.9\linewidth]{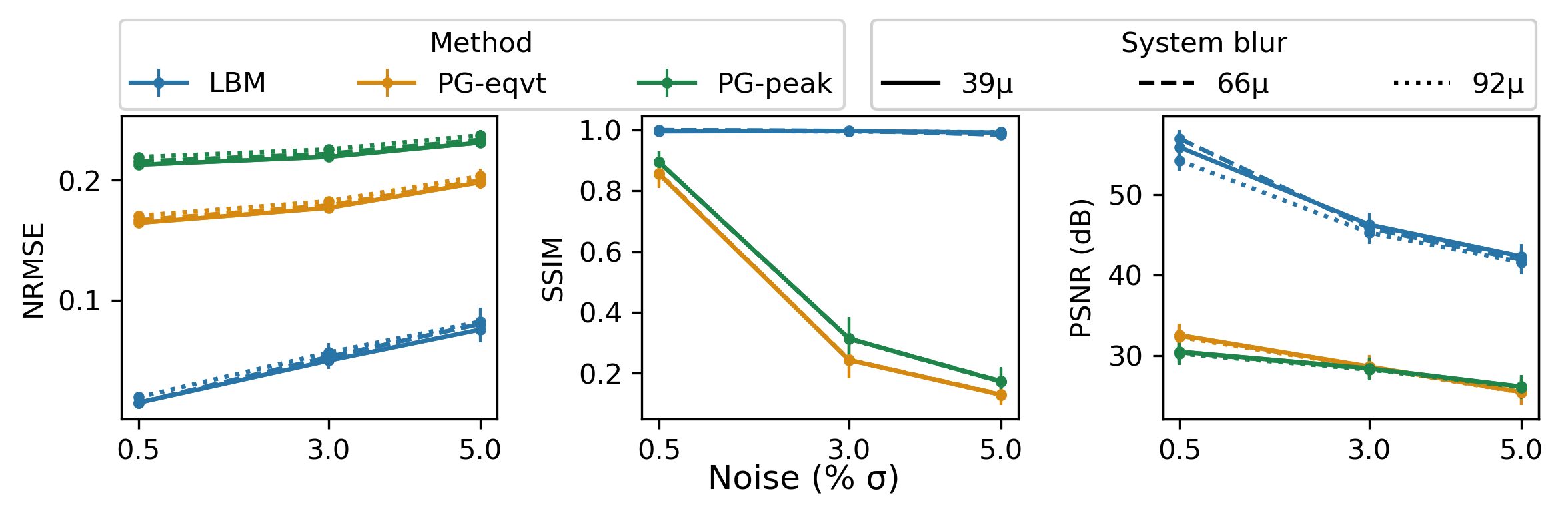} 
\end{subfigure}
\caption{Results for variations in blur and noise. The LBM outperforms conventional methods in all three metrics, particularly PSNR and SSIM at high noise levels. This indicates better retention of qualitative features and denoising effects due to the LBM as compared to the conventional methods. Furthermore, an increase in noise levels from 0.5\% to 5\% had a much larger effect on performance than variation in blur in the range 39 $\mu$m to 92 $\mu$m FWHM, as seen in the near overlap of values of all performance metrics on varying blur at each noise level.}
\label{fig2}
\end{figure}

Quantitative results from the first study, wherein noise and system blur were varied to study the combined effects of these physical factors on the performance of the LBM demonstrate that although the NRMSE for the LBM was $<10\%$ in all cases, increasing system noise and system blur levels resulted in degraded quantitative performance (see Figure \ref{fig2}) as expected. However, the  NRMSE was more impacted (6.5\% vs 0.7\%) by an increase in system noise (from 0.5\% to 5\%) than by increase in system blur (from 39 $\mu$m to 92 $\mu$m), at intermediate values of the other parameter. Similar trends were observed for the SSIM and PSNR (see Figure \ref{fig2}), indicating the superior quantitative performance of the LBM, especially at high levels of noise. This revealed that, although blur and noise worsened the performance of the LBM, it was stable within typical ranges of these system parameters found in laboratory conditions, and has better quantitative performance than conventional methods even at high levels of noise.

\begin{figure}[h!bt]
\centering
\begin{subfigure}{0.9\textwidth}
    \includegraphics[width=0.9\linewidth]{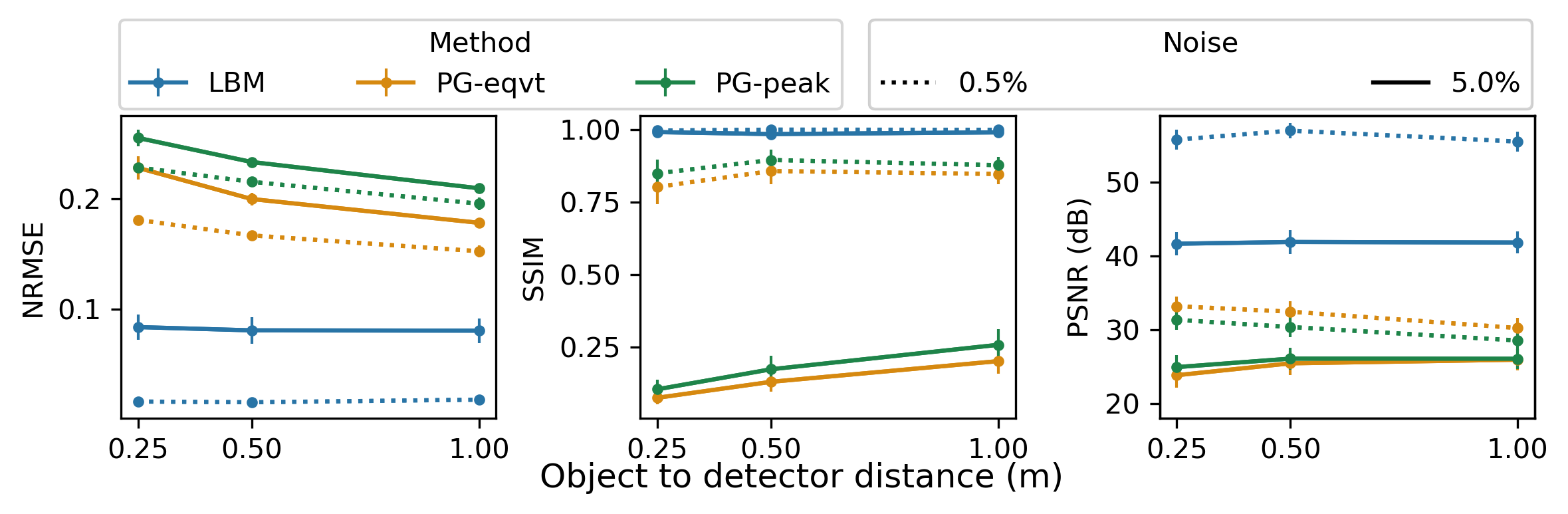} 
\end{subfigure}
\caption{Results for variation in object-to-detector distance ($R_2$) at two noise levels. Across all three metrics, the performance of the LBM was better than the conventional methods. Varying $R_2$ resulted in negligible variation in the performance of the LBM, while higher noise levels resulted in worse NRMSE and PSNR, but similar SSIM. This indicates good qualitative recovery of features but slightly worse quantitative phase recovery at high noise levels.}
\label{fig3}
\end{figure}

Similarly, to study the impact of varying the object-to-detector distance, three choices of $R_2$: 0.25 m, 0.5 m to 1 m ($R_1$=1 m) were considered, corresponding to magnifications of 1.25, 1.5 and 2 respectively. It was observed that the change in the quantitative performance (NRMSE) was negligible for the LBM across all $R_2$ values, and the NRMSE was consistently lower than the corresponding values of the conventional methods (Figure \ref{fig3}). Note that the effects of magnification and propagation distances are related, and the LBM is expected to account for both effects simultaneously. Moreover, the impact of variation in noise (0.5\% to 5\%) on quantitative performance of the LBM was greater ($\approx6.5\%$ vs $\approx0.3\%$) than the impact of variations in $R_2$ (from 0.25 m to 1 m). 

\subsection{Behaviour of the network under variations in materials and phantoms}
As the distribution of refractive indices varies with materials and energy, it is important to assess the extent to which the LBM can learn how to recover the phase when the object is heterogeneous. Hence, to study the performance of the LBM in the presence of material heterogeneity, the following cases were considered: (Case 1) multiple materials with no embedding medium, and (Case 2) multiple materials in an embedding medium; these correspond to configurations (b) and (c) of the spheres phantom (see Sec.3.2). In these studies, the system settings were fixed at $R_1$=1 m, $R_2$=0.5 m, blur=66 $\mu$m FWHM and noise=5\%. In the two cases, results from the LBM were compared against the conventional method PG (for Case 1) and BT (for Case 2), employed separately for each material, followed by splicing.

\begin{figure}[h!tb]
\centering
\begin{subfigure}{0.8\textwidth}
    \includegraphics[width=0.9\linewidth]{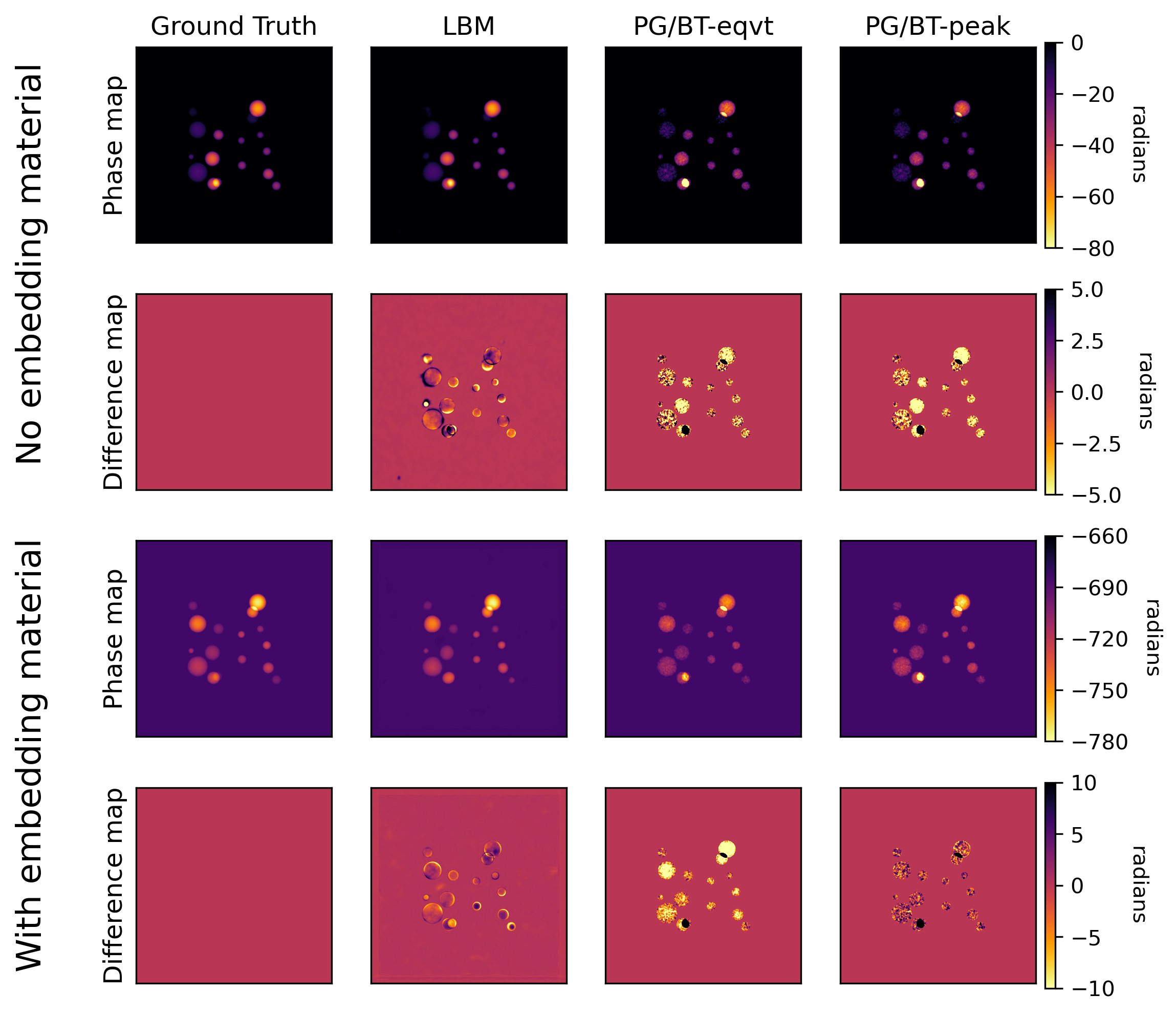} 
\end{subfigure}
\caption{Qualitative results for QPR from a multi-material phantom: in the absence of an embedding medium (Case (1)), and in its presence (Case (2)). Despite exhibiting well-formed features (see rows 1 and 3), the conventional methods show strong errors as seen in the difference maps (rows 2 and 4). The LBM outperformed the conventional methods especially in regions with overlapping structures. Furthermore, predictions from the conventional methods are clearly noisier than predictions from LBM. Also, recall that BT requires the knowledge of all object locations for performing QPR, unlike the LBM.}
\label{fig5}
\end{figure}

Qualitative results from the multi-material phantom study (Figure\ref{fig5}) demonstrate that although phase predictions from all methods exhibited well-formed features, the LBM clearly outperformed the conventional methods in regions of overlapping structures and in terms of denoising performance.

Quantitative results from both cases of the multi-material phantom study show that the LBM outperforms PG and BT \citep{beltran20102d} in terms of pixel-wise accuracy (NRMSE) and PSNR (see Table \ref{table1}). Although the SSIM was similar ($>=$0.96) for the LBM and the conventional methods, quantitative errors were clearly greater for the conventional methods than the LBM as seen in the NRMSE ($>$30\% in Case 1 and $>$0.2\% in Case 2) and PSNR ($>$20dB). Note that the NRMSE appeared to be much lower in the presence of an embedding medium, i.e., Case (2). This is because it is a measure of relative error, and contains the effects of a larger background phase value ($\approx680$ radians) as compared to that in Case (1) (0 radians).
Quantitative and qualitative results from the multi-material phantom, together indicate that the LBM could successfully perform QPR by recovering a non-linear mapping between phase and intensity, and not simply a linear mapping for a single ``effective" material.

\begin{table}[h!tb]
\begin{center}
\begin{tabular}{ccccc}
    \hline
    \multirow{2}{*}{Multi-material object cases}& \multirow{2}{*}{Measure of quality}& \multicolumn{3}{c}{Methods}  \\
    & & LBM & PG/BT-eqvt & PG/BT-peak\\
    \hline
    \multirow{3}{*}{Embedding medium absent} & NRMSE (\%) & \textbf{9$\pm$1} & $40\pm13$ & $37\pm11$\\
    & SSIM & \textbf{0.98} & $0.95$ & $0.96$ \\
    & PSNR (dB) & \textbf{41$\pm$2} & $26\pm2$ & $26\pm2$ \\
    \hline
    \multirow{3}{*}{Embedding medium present} & NRMSE(\%) & \textbf{0.5$\pm$0.04} &$0.7\pm0.2$ & $0.9\pm0.4$\\
    & SSIM & \textbf{0.99} & $0.96$ & $0.96$ \\
    & PSNR (dB) & \textbf{41$\pm$2} & $27\pm1$ & $25\pm3$ \\
    \hline
\end{tabular}
\caption{Quantitative results for multi-material phantoms in the absence of an embedding medium: Case (1), and in its presence: Case (2). For a multi-material phantom, irrespective of the presence or absence of an embedding medium, the LBM clearly outperforms PG and BT in terms of NRMSE and PSNR, while demonstrating performance similar to PG/BT methods in terms of the SSIM.}
\label{table1}
\end{center}

\end{table}

Finally, the effect of foreground area was studied to determine whether increasing the information content per realization, i.e., whether a greater number of pixels having a value other than that of the constant background within a realization aided quantitative performance. This is important to understand because the MSE loss function was employed for training. The system settings were: $R_1$=1 m, $R_2$=0.5 m, blur=66 $\mu$m FWHM and noise=5\%. It was found that foreground area variations in the range 10-30\% did not cause any significant change in the quantitative performance of the network ($<0.5\%$), indicating that the training ensemble of images produced from the designed phantoms contained sufficient instances of intensity-phase mappings to learn QPR.

\subsection{Generalization studies}
Generalization studies were performed to investigate the robustness of a pre-trained LBM when applied to data that differed in some characteristics from the data employed in model training. In the machine learning literature, such data are referred to as ``out-of-distribution" data (OOD).  In the studies below, OOD data were produced by considering variations in (i) propagation distances and (ii) phantom structure. Experimental data were also employed as a type of OOD data.
\subsubsection{Robustness to propagation distances}
\begin{figure}[h!tb]
\centering
\begin{subfigure}{0.9\textwidth}
    \includegraphics[width=0.9\linewidth]{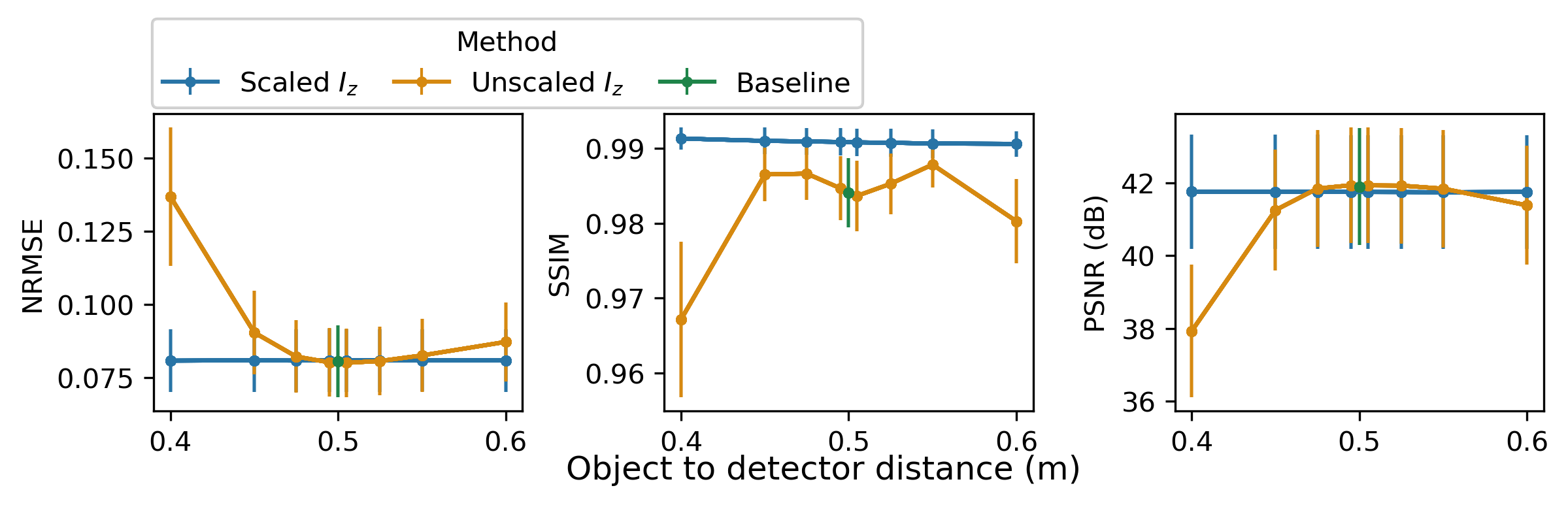} 
\end{subfigure}
\caption{Generalization performance on varying propagation distance for a network trained with data corresponding to $R_2$=0.5 m and tested on datasets with $R_2$ ranging from 0.4 to 0.6 m. Scaling $I_z$ by the correct flatfield, i.e., corresponding to $R_2$(test) improves the robustness of the network for all three metrics as compared to the unscaled dataset, i.e., corresponding to $R_2$(train). This effect is more pronounced at lower values of $R_2$.}
\label{fig7}
\end{figure}

A LBM trained with data corresponding to a fixed set of system parameters ($R_1$=1 m, $R_2$=0.5 m, blur=66 $\mu$m FWHM and noise=5\%) was employed to perform phase retrieval by use of OOD test datasets produced by introducing variations in propagation distances ($R_2$=0.4 -- 0.6 m). As the flatfield intensity is also impacted by variations in the propagation distance, additional error (besides error due to OOD measurements) maybe incurred due to incorrect flatfield normalization. Hence, two cases of normalization were considered for the OOD measurements $I_{z\in OOD}$, where the subscript indicates that the propagation distance $z$ was OOD (and did not correspond to training measurements $I_{z=0.5 m}$). The two cases of normalization were: (i) Unscaled $I_z$, where $I_{z\in OOD}$ measurements were flatfield corrected by the use of flatfield measured at $I_{z=0.5m}$ and (ii) Scaled $I_z$, where $I_{z\in OOD}$ measurements were flatfield corrected appropriately by the use of corresponding flatfield (acquired at the same OOD value of $z$). As seen in Figure \ref{fig7}, the pre-trained network was robust for variations of at least 20\% in the propagation distance when $I_z$ was scaled according to Case (ii) with negligible change in quantitative performance. This was different from Case (i), which resulted in an increase of up to 7.5\% in NRMSE for the same range of $R_2$. Qualitatively, although features were well formed, a slight asymmetry was also observed occasionally; this might be due to the mismatch in magnification between the training and testing datasets. Thus, the LBM, with proper normalization of data, might have sufficient robustness for application under experimental settings, where the accuracy of the measured propagation distance might be inexact due to practical considerations. 

\subsubsection{Robustness to object structure}
\begin{figure}[h!tb]
\centering
    \includegraphics[width=0.6\linewidth]{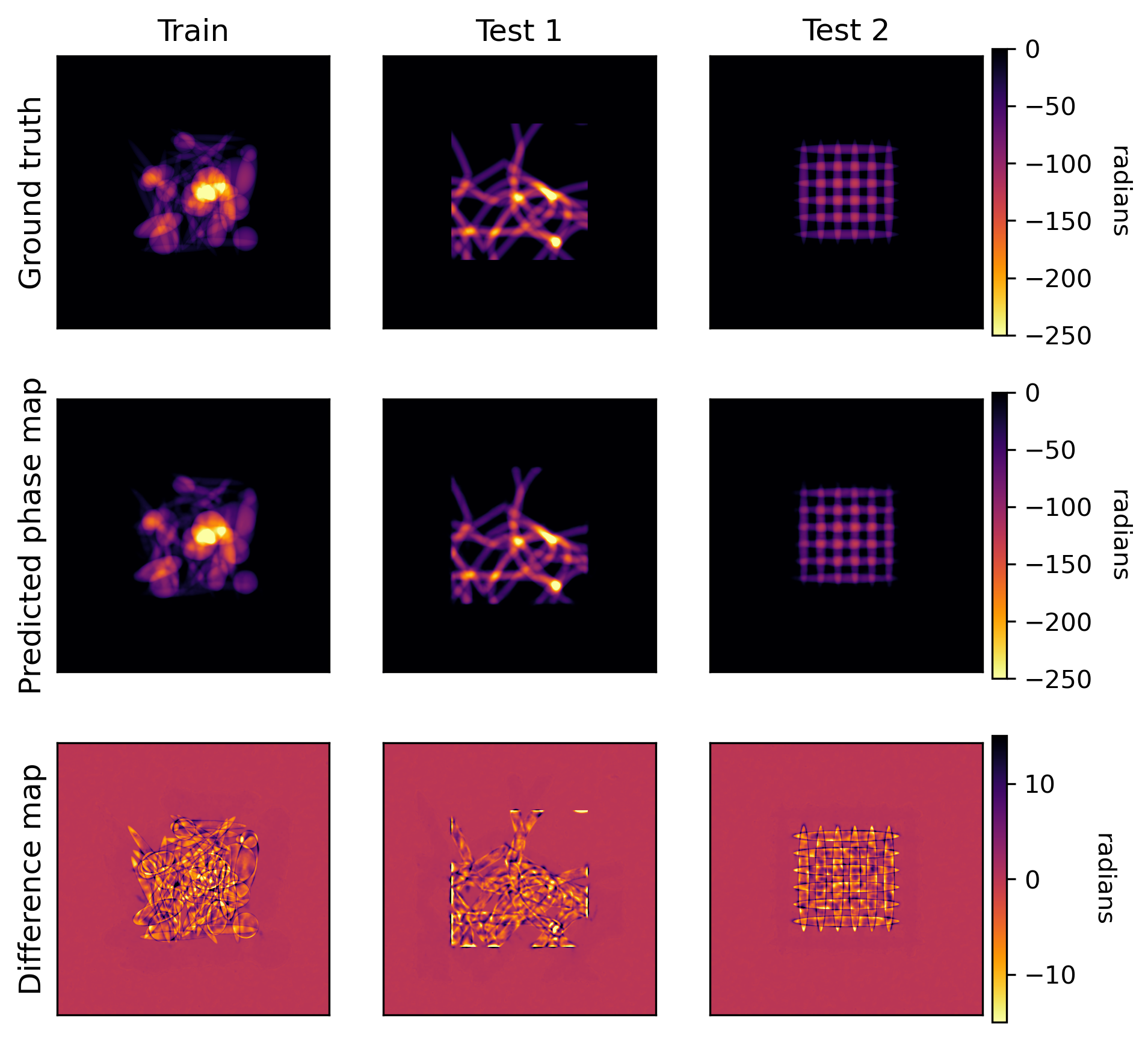} 
\caption{Generalization results for object structure. A network trained on complex phantom realizations (top left) generalizes well to out-of-distribution test examples (top center and right). Quantitative and qualitative performance for the test cases is at par with the training dataset (NRMSE $<$ 10\%). Note that only object structure is varied in this study, the object material and system geometry is the same across training and test examples.}
\label{fig8}
\end{figure}

To test the robustness to variations in object structure, a LBM trained on the complex phantom (described in Sec. 3.2) was employed for phase retrieval on test data that corresponded to structurally distinct objects imaged under the same system settings ($R_1$=1 m, $R_2$=0.5 m, blur=66 $\mu$m FWHM and noise=5\%) and for the same material allocations (object material: soft-tissue). The test data consisted of two examples: (i) an irregular structure consisting of multiple, vessel-like sub-structures \citep{beutel2000handbook} (mean radii: 40, 8 pixels), and (ii) a regular structure constituting of equidistant ellipsoids (semi-axes lengths: 20, 200, 50 pixels) placed on a grid. It was observed that the network generalized well to variations in object structure (Figure \ref{fig8}). This revealed that a LBM is not constrained by accurate structural modeling of the test object. This, in turn, implies that the LBM could be used for inference from complex object measurements even when trained with measurements produced by imaging a collection of structurally simpler objects. 

\subsubsection{Generalizability to experimental data}
Polystyrene spheres of radius 1.94$\pm$0.05mm were experimentally imaged to acquire two images: a contact plane image ($R_1$ = 3.16 m, $R_2$ = 0.02 m) and a downstream measured image ($R_1$ = 1.24 m, $R_2$ = 1.94 m, FOV = 20 mm). The focal spot size was 12 $\mu$m, while all other system parameters were as described before (in Sec. 3.3). The OOD test dataset compromised 15 measurements, each of which corresponded to an image-pair acquired by imaging a physical phantom consisting of 5 polystyrene spheres. A network trained on simulated data corresponding to the spheres phantom, but matched in material, distribution of radii, and system geometry, was employed for predictions on experimental data. Prior to prediction, the experimental data were flatfield corrected, intensity outliers were removed by saturating pixels in the extreme 0.5 percentiles, and the data were normalized as per the training dataset. Quantitative phase predictions produced by the LBM yielded superior accuracy in maximum depth (94$\pm$6\%), as opposed to predictions from PG-eqvt (66$\pm$1\%) and PG-peak (86$\pm$1\%). The LBM performed comparably in terms of equivalent diameter (97$\pm$2\% vs 99$\pm$1\% for both PG methods) and roundness (90$\pm$1\% vs 90$\pm$7\% for both PG methods). Note that these results for the LBM indicate a \textit{baseline} performance in the absence of any transfer learning, which may potentially improve the generalizability to experimental data by addressing certain inaccuracies in system modeling. 

\begin{figure}[htb]
\centering

\begin{subfigure}{0.4\textwidth}
    \includegraphics[width=0.9\linewidth]{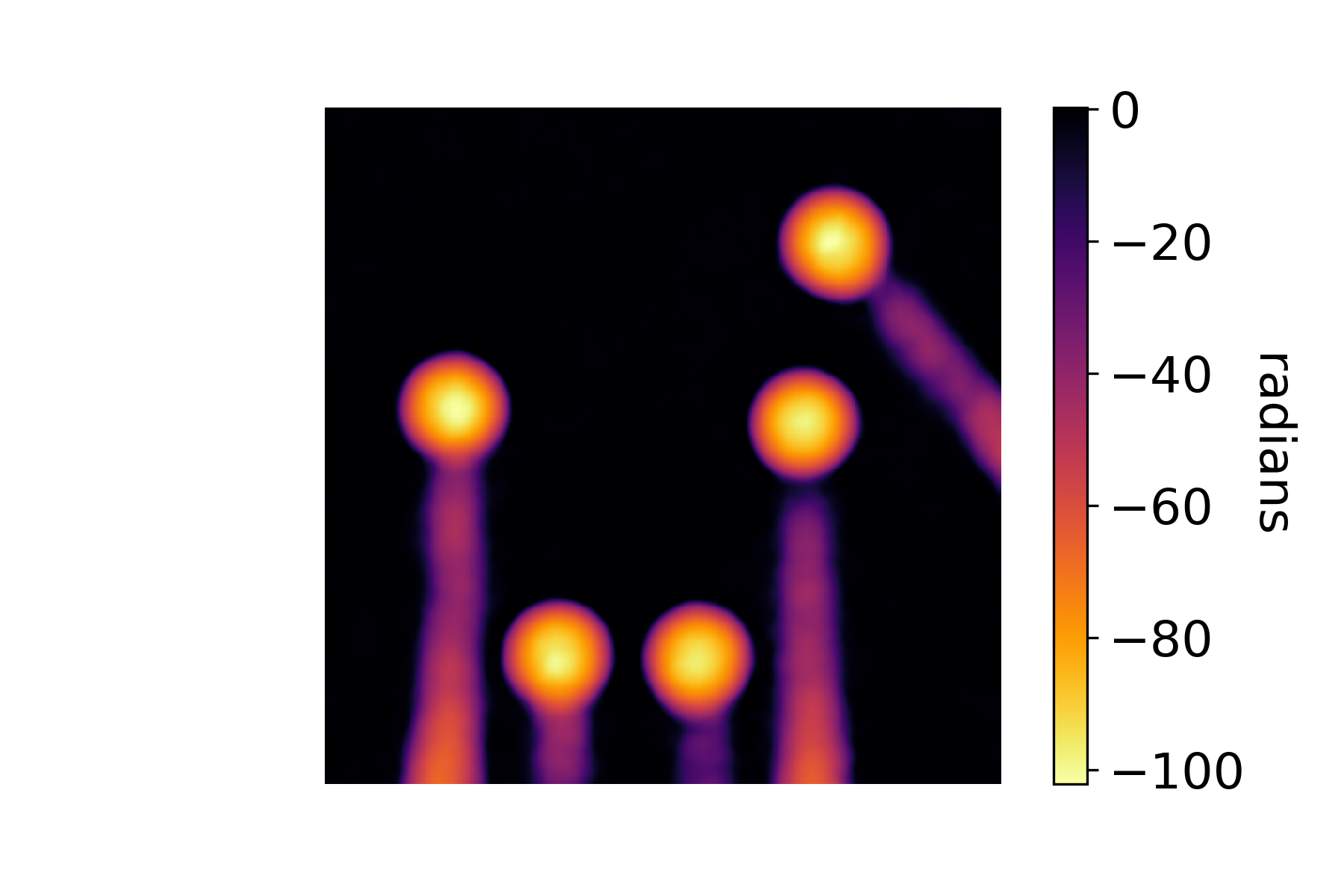} 
\end{subfigure}
\hspace*{-3em}
\begin{subfigure}{0.4\textwidth}
    \includegraphics[width=0.9\linewidth]{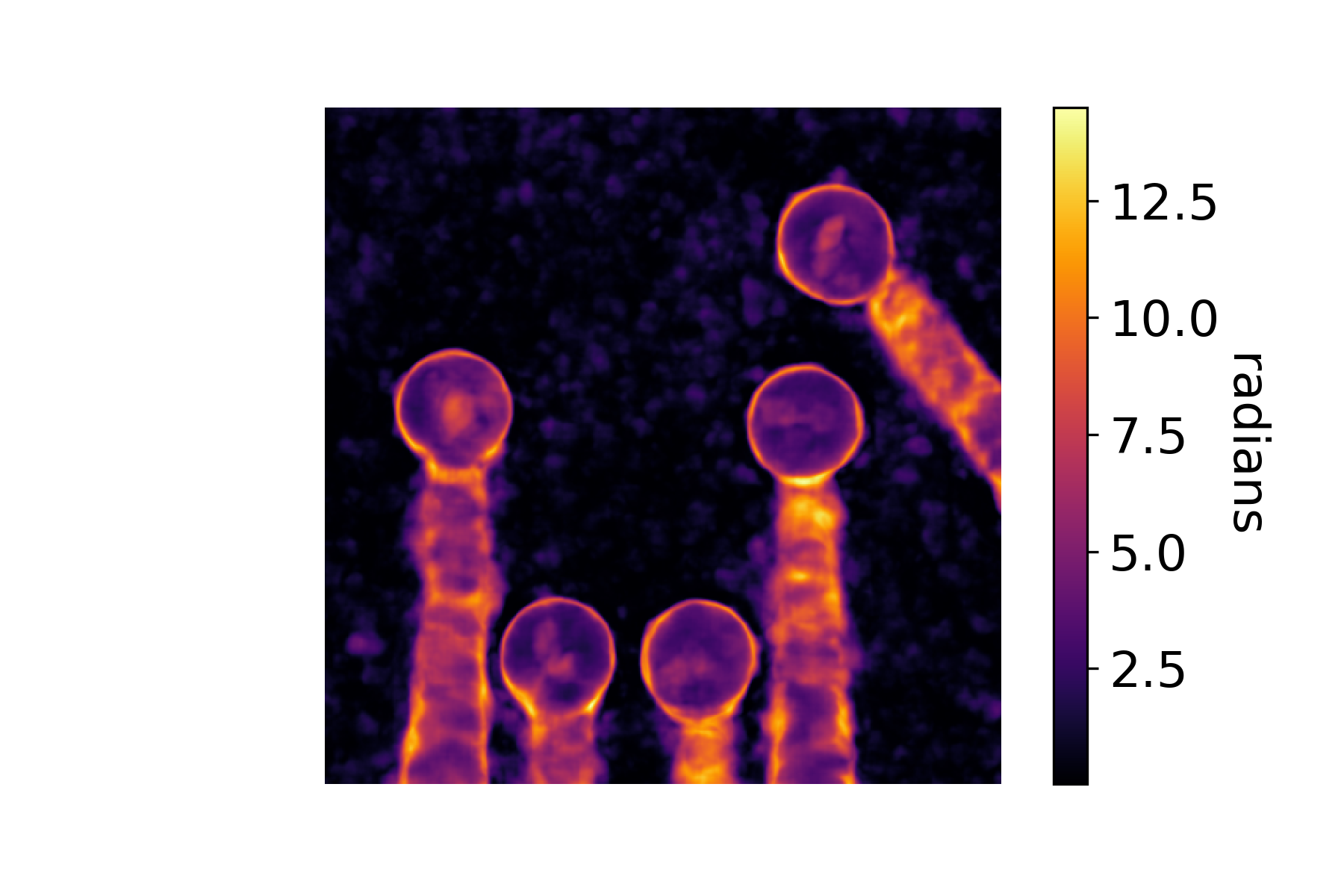} 
\end{subfigure}

\caption{Mean and standard deviation computed over 15 predictions of phase from experimental images of polystyrene spheres. Qualitatively, results from experimental data demonstrate generally well formed structures in the mean prediction (left) and some instances of inadequate denoising in the standard deviation image (right). The true value of phase at the center of each projected sphere is -96$\pm$2 radians.}
\label{fig9}
\end{figure}

Qualitative results (mean and standard deviation images over 15 predictions) are shown in Figure \ref{fig9}. The spheres appear well-formed and noise is largely suppressed. The background does show instances of slightly inadequate noise suppression, possibly due to the mismatch in the simulated noise distribution and the experimental system noise; however, noise suppression is far superior to the PG methods. Thus, the LBM outperformed PG in terms of noise suppression and estimation of maximum depth, while performing comparably in terms of qualitative aspects as measured by equivalent diameter and roundness. This indicates that an LBM trained on simulated data can generalize well to experimental data when system modeling is accurate.  

\section{Discussion}
\subsection{Behavior under system and object variations}
The LBM is stable in quantitative performance under various conditions of system blur, noise and object-to-detector distances as seen in Figs. \ref{fig2} and \ref{fig3}. In addition, the qualitative results demonstrate a remarkably superior denoising effect than the conventional methods on visual comparison. Together, the stability in quantitative and qualitative performance over typical system settings suggests that an LBM is applicabile  over a wide range of typical system conditions, and without strong limiting assumptions about the object. Most importantly, results from the generalization studies demonstrate the robustness of the LBM to minor variations in propagation distances typical to practical conditions, where the precision of acquiring images at a given propagation distance might be low due to limitations of instrumentation. Furthermore, the robustness to changes in object structure can prove beneficial in the absence of accurate models of object or large experimental datasets. Last, results from the experimental studies demonstrate the applicability of this method to experimental laboratory conditions, provided that the system modeling is accurate. Further improvement in experimental studies potentially could be achieved via explicit modeling of additional system parameters or implicitly via transfer learning methods. Thus, this work demonstrated the potential utility of learning-based approaches for QPR in typical scenarios encountered in experimental PB-XPC imaging under laboratory conditions. 

\subsection{Advantages and disadvantages of a LBM}
An LBM enables the learning of a non-linear mapping without strong restrictions on the composition of the object, unlike many analytic methods for QPR \citep{langer2008quantitative}. In addition, quantitative phase maps for multi-material objects can be retrieved at once, without pair-wise consideration of materials \citep{haggmark2017comparison} or post-hoc splicing. However, the accuracy of such an end-to-end LBM is data dependent: it is affected by the accuracy in system modeling as well as by sufficiency in representation of the underlying mapping of the heterogeneous object to the phase map. Note that representational sufficiency is with regard to material overlaps only and \textit{not} object structures. Thus, an LBM trained on a phantom consisting of specific materials could translate well to a structurally distinct phantom consisting of the same materials, as demonstrated in the object structure generalization study. There remains a need to compare end-to-end LBMs, as employed in this study, against other LBMs that incorporate system knowledge to investigate improvements in generalizability of a learning-based approach.  

\subsection{Plausibility of a convolutional neural network- (CNN) based learning method as a solution for QPR}
Phase contrast in intensity images manifests as the fringes present at the material edges in an object. The characteristic size of each fringe depends on the following system considerations: system blur, propagation distance, and wavelength, besides object attenuation \citep{nesterets2005optimization, gureyev2008some}. Thus, in order to retrieve phase from polychromatic data under laboratory conditions, two operations have to be concurrently performed: (i) implicit filtering or extraction of the dominant, spatio-temporally coherent mode \citep{born2013principles, zysk2010transport} from the input intensity measurements and, (ii) phase retrieval corresponding to this extracted dominant mode. The first problem may be represented as a deconvolution and denoising problem; the chosen LBM architecture has been employed as a solution for the two problems in other imaging scenarios \citep{park2018computed, reymann2019u, lee2020mu}. The second problem, QPR from the dominant coherent mode, is an ill-posed, inverse problem. Solution to this problem, in the context of the TIE, requires knowledge of the transverse derivatives (gradient and Laplacian) of the two input intensity images. These operations, too, can be performed by the convolutional neural network (CNN)- based architecture of the LBM. Thus, a CNN-based method can prove to be a plausible solution to the polychromatic phase retrieval problem.

\subsection{Training considerations}
A LBM aims to learn the mapping between phase and intensity measurements, independent of the phantom structure and design. In order to retain this mapping during data pre-processing, while simultaneously ensuring an appropriate range of data for gradient computation, it is essential to normalize over the entire stack of training images and not over individual slices. Per-slice normalization would render the problem unsolvable. Furthermore, the validation and test data sets are also mapped onto the same, normalized scale. Thus, retention of this inverse mapping has to be ensured not just during pre-processing and training, but also during inference.

As the pixel-wise MSE loss was employed for training,  quantitative performance was prioritized over qualitative feature information. Alternatively, if the goal is excellent qualitative performance alone, alternative loss functions that take into account the correlations in object structure such as normalized correlation co-efficient (NCC) \citep{goy2018low}, perceptual loss \citep{deng2020probing} or even CNR, could be employed to focus on qualitative feature information. To this end, a variation in network architecture such as mixed-scale dense networks \citep{pelt2018mixed, mom2022mixed} instead of convolutional networks could be employed. Another effect of employing the MSE loss is the strong denoising performance, given the large proportion of pixels forming the background. However, the superior denoising performance might be partially offset by slightly worse performance for low phase contrast-inducing materials present within multi-material objects, and under high noise conditions, as the network training minimizes MSE and prioritizes superior performance on high phase contrast-inducing materials. An alternative training strategy might include material-specific weights in the loss function, or involve training specifically at various quantitative ranges of phase. Furthermore, with a focus on QPR and employment of a pixel-wise loss, high frequency artifacts were observed at the edges, especially as residues at the locations of the magnified objects. These artifacts were quantitatively minimal, and could be removed via post-processing. Before the practical deployment of any LBM, it is important to identify, a priori, the kind of characteristic artifacts that may be produced by the LBM as opposed to artifacts from conventional methods that can be determined given their analytical formulation \citep{rodgers2020optimizing}. This could inform a post-processing method that might be employed to eliminate the same.

Last, in case of multi-material objects in an embedding medium, the constant background phase and phase variations relative to the background could span different orders of magnitude. With MSE as the loss function, learning occurs at two scales - (i) estimating the background, which often consists of more pixels than the foreground and thus, could provide large improvement in MSE if correctly estimated, and (ii) estimating the variations in phase with respect to the background. In order to improve training performance in this scenario, an adaptive learning rate, or offsetting the constant background value was found to be useful. 

\section{Conclusion}
In conclusion, we employed a LBM and assessed its robustness as a solution for QPR from polychromatic intensity measurements for PB-XPC imaging under laboratory conditions. It was found that the method is stable under typical variations in system settings and can be applicable for complex, multi-material objects. Last, the method generalized well to previously unseen, complex phantoms as well as to minor variations in propagation distances and, hence, has potential applicability for QPR from measurements of complex objects imaged experimentally via PB-XPC under laboratory conditions.

\section{Acknowledgement}
This work made use of the Illinois Campus Cluster, a computing resource that is operated by the Illinois Campus Cluster Program (ICCP) in conjunction with the National Center for Supercomputing Applications (NCSA) and which is supported by funds from the University of Illinois at Urbana-Champaign. This research is also part of the Delta research computing project, which is supported by the National Science Foundation (award OCI 2005572), and the State of Illinois. Delta is a joint effort of the University of Illinois at Urbana-Champaign and its National Center for Supercomputing Applications. This work was supported in part by NIH awards R01 EB018525 and R01 EB020604. RD acknowledges support from the Imaging Sciences Pathway Trainee Fellowship (T32EB01485505). We thank Varun A. Kelkar for sharing image data employed in the simulation of test example (i) described in Sec. 4.3.2.
\newcommand{\newblock}{}
\bibliographystyle{dcu-italics}
\bibliography{references}
\end{document}